\documentclass[12pt]{article}

\newtheorem{fact}{Fact}
\newtheorem{remark}{Remark}
\newtheorem{theorem}{Theorem}[section]
\newtheorem{corollary}[theorem]{Corollary}
\newtheorem{lemma}[theorem]{Lemma}
\newtheorem{claim}[theorem]{Claim}
\newtheorem{proposition}[theorem]{Proposition}
\newtheorem{property}[theorem]{Property}

\newtheorem{definition}[theorem]{Definition}
\newtheorem{observation}[theorem]{Observation}
\newtheorem{example}[theorem]{Example}

\newcommand{\proof}{{\bf Proof.}}

\usepackage{epsfig}
\usepackage{amsmath}

\usepackage{amsfonts}

\def\nottoobig#1{{\hbox{$\left#1\vcenter to1.111\ht\strutbox{}\right.\n@space$}}}

\if01

\newtheorem{theorem}{Theorem}[section]

\newtheorem{proposition}[theorem]{Proposition}

\fi

\newcommand{\ie}{$\mbox{i.e.}$}

\newlength{\filength}
\newsavebox{\gcbox}
\sbox{\gcbox}{\framebox[\filength]{\rule{0ex}{2ex}}}

\newcommand{\qedblob}{\mbox{\rule[-1.5pt]{5pt}{10.5pt}}}
\def\literalqed{{\ \nolinebreak\hfill\mbox{\qedblob\quad}}}

\def\qed{\literalqed}

\newcommand{\singlespacing}{\let\CS=
\@currsize\renewcommand{\baselinestretch}{1}\tiny\CS}
\newcommand{\singlespacingplus}{\let\CS=
\@currsize\renewcommand{\baselinestretch}{1.25}\tiny\CS}
\newcommand{\doublespacing}{\let\CS=
\@currsize\renewcommand{\baselinestretch}{1.75}\tiny\CS}
\newcommand{\draftspacing}{\let\CS=
\@currsize\renewcommand{\baselinestretch}{2.0}\tiny\CS}

\def\zo{\{0,1\}}

\def\mapping{\rightarrow}

\newcommand{\prob}{\rm Prob}

\newcommand{\poly}{{\rm poly}}

\makeatletter
\def\@listI{\leftmargin\leftmargini \parsep 4.5pt plus 1pt minus 1pt\topsep6pt plus 2pt minus 2pt \itemsep  2pt plus 2pt minus 1pt}

\let\@listi\@listI
\@listi
\makeatother

\author{ {Marius Zimand\/}
\thanks{  Department of Computer and Information Sciences, Towson University,
Baltimore, MD.; email: mzimand@towson.edu; http://triton.towson.edu/\~{ }mzimand.
This work has been supported
by NSF grant CCF 1016158.}}

\begin{document}
\title{On efficient constructions of short lists containing mostly Ramsey graphs}

\date{}

\maketitle

\begin{abstract}
One of the earliest and best-known application of the probabilistic method is the proof of existence of a $2 \log n$-Ramsey graph, \ie, a graph with $n$ nodes that contains no clique or independent set of size $2 \log n$. The explicit construction of such a graph is a major open problem. We show that a reasonable hardness assumption implies that in polynomial time one can construct a list containing polylog($n$) graphs such that most of them are $2 \log n$-Ramsey.
\end{abstract}

\section{Introduction}

A $k$-Ramsey graph is a graph $G$  that has no clique of size $k$ and no independent set of size $k$. It is known that for all sufficiently large $n$, there exists a $2 \log n$-Ramsey graph with $n$ vertices. The proof is nonconstructive, but of course such a graph can be built in exponential time by exhaustive search. A major line of research is dedicated to constructing a $k$-Ramsey graph with $n$ vertices with $k$ as small as possible and in time that is bounded by a small function in $n$, for example in polynomial time, or in quasi-polynomial time, DTIME[$2^{{\rm polylog}(n)}$]. Till recently, the best polynomial-time construction of a $k$-Ramsey graph with $n$ vertices has been the one by Frankl and Wilson~\cite{frawil:j:ramsey}, for $k = 2^{\widetilde{O}(\sqrt{\log n})}$. Using deep results from additive combinatorics and the theory of randomness extractors and dispersers, Barak, Rao, Shaltiel and Wigderson~\cite{barraoshawig:c:ramseygraph} improved this to $k=2^{(\log n)^{o(1)}}$. Notice that this is still far off from $k = 2 \log n$.

As usual when dealing with very difficult problems, it is natural to consider easier versions. In this case, one would like to see if it is possible to efficiently construct a small list of $n$-vertices graphs with the guarantee that one of them is $2 \log n$-Ramsey. The following positive results hold.
\begin{theorem}
\label{t:ramseyuncond} There exists a quasipolynomial-time algorithm that on input $1^n$ returns a list with $2^{\log^3 n}$ graphs with $n$ vertices, and most of them are $2 \log n$-Ramsey. In fact, since in quasipolynomial time one can check whether a graph is $2 \log n$-Ramsey, the algorithm can be modified to return one graph that is $2 \log n$-Ramsey.
\end{theorem}
\begin{theorem}
\label{t:ramseycond}  Under a reasonable hardness assumption $H$, there exists a constant $c$ and a polynomial-time algorithm
that on input $1^n$ returns a list with ${\log^c n}$ graphs with $n$ vertices, and most of them are $2 \log n$-Ramsey. 
\end{theorem}
The proofs of these two results use basic off-the-shelf derandomization techniques. The proof (one of them) of Theorem~\ref{t:ramseyuncond} notices that the probabilistic argument that shows the existence of $2 \log n$-Ramsey graphs only needs a distribution on the set of $n$-vertices graphs that is $2 \log^2 n$-wise independent. There exist such distributions whose support have the following properties: (a) the size is $2^{O(\log^3 n)}$ and (b) it can be indexed by strings of size $O( \log^3 n)$. Therefore if we make an exhaustive search among these indeces, we obtain the result.

 Theorem~\ref{t:ramseycond} uses a pseudo-random generator $g$ that can fool NP-predicates. The assumption $H$, which states that there exists a function in $E$ that, for some $\epsilon >0$, requires circuits with SAT gates of size $2^{\epsilon n}$, implies the existence of such pseudo-random generators. Then going back to the previous proof, it can be observed that the property that an index corresponds to a graph that is not $2 \log n$-Ramsey is an NP predicate. Since most indeces correspond to graphs that are $2 \log n$-Ramsey, it follows that for most seeds $s$, $g(s)$ is also $2 \log n$-Ramsey. Therefore, it suffices to make an exhaustive search among all possible seeds. Since a seed has length $O( \log |\mbox{index}|) = O(\log \log^3 n)$, the result follows.

Theorem~\ref{t:ramseycond} can be strengthened to produce a list of concise representations of graphs. A string $t$ is a \emph{concise representation} of a graph $G = (V,E)$ with $V = \{1, \ldots, n\}$  if there is an algorithm $A$ running in time $\poly (\log n)$ such that for every $u \in V, v \in V$, $A(t,u,v) = 1$ if $(u,v) \in E$ and $A(t, u,v)=0$ if $(u,v) \not\in E$. With basically the same proof as that of Theorem~\ref{t:ramseycond}  one can show the following result.

\begin{theorem}
\label{t:ramseystrongcond}
Under a reasonable hardness assumption $H$, there exists a constant $c$ and an algorithm running in time $\poly(\log n)$ that on input $n$ (written in binary notation) returns a list $t_1, \ldots, t_{\log^c n}$, and most elements of the list are concise representations of $2 \log n$-Ramsey graphs.
\end{theorem}

Theorem~\ref{t:ramseyuncond} is folklore. It appears implicitely in the paper of M. Naor~\cite{mnaor:c:ramsey}. Theorem~\ref{t:ramseycond} may also be known, but we are not aware of any published statement of it. Fortnow in the \emph{Computational Complexity} blog~\cite{for:t:derandramsey} and Santhanam at the 2011 Bertinoro seminar on Ramsey theory~\cite{san:t:ramsey} mention a weaker version of Theorem~\ref{t:ramseycond}, in which the same hardness assumption is used but the size of the list is polynomial instead of polylogarithmic. This motivated us to write this note.

Section~\ref{s:remarks} contains some additional remarks. First we analyze the implication of Theorem~\ref{t:ramseycond} when plugged in a construction of M. Naor~\cite{mnaor:c:ramsey} that builds a $k$-Ramsey graph from a list of graphs, most of which are $k'$-Ramsey graphs, which is exactly what Theorem~\ref{t:ramseycond} delivers. We notice that the parameters obtained in this way are inferior
to the result of Barak et al.~\cite{barraoshawig:c:ramseygraph}. Secondly,  we consider the problem of explicit lower bounds for the van der Waerden Theorem, a problem which is related to the explicit construction of Ramsey graphs. We notice that the hardness assumption which derandomizes BPP implies lower bounds for the van der Waerden Theorem  that match the non-constructive lower bounds obtained via the Lovasz Local Lemma. The original proof of the Lovasz Local Lemma does not seem to yield this result. Instead we use a proof of Gasarch and Haeupler~\cite{gas-hae:j:waerden},  based on the methods of Moser~\cite{mos:c:lovasz} and Moser and Tardos~\cite{mostar:j:lovasz}. 
\section{The  hardness assumption}

The hardness assumption needed in theorem~\ref{t:ramseycond} is that there exists a function $f$ computable in $E$ (where 
$E = \bigcup_{c} {\rm DTIME}[2^{cn}]$) that, for some $\epsilon >0$, cannot be computed by circuits of size $2^{\epsilon n}$ that also have SAT gates (in addition to the standard logical gates).  More formally let us denote by $C_f^{{\rm SAT}}(n)$ the size of the smallest circuit with SAT gates that computes the function $f$ for inputs of length $n$. 
\medskip

Assumption $H$:
There exists a function $f$ in $E$ such that, for some $\epsilon > 0$, for every $n$, $C_f^{{\rm SAT}}(n) > 2^{\epsilon n}$.
\medskip

Klivans and van Melkebeek~\cite{km:j:prgenoracle}, generalizing the work of Nisan and Wigderson~\cite{nis-wig:j:hard} and Impagliazzo and Wigderson~\cite{imp-wig:c:pbpp}, have shown that, under assumption $H$, for every $k$, there is a constant $c$ and a pseudo-random generator $g:\zo^{c \log n} \mapping \zo^n$, computable in time polynomial in $n$, that fools all $n^k$-size circuits with SAT gates. Formally, for every circuit $C$ with SAT gates of size $n^k$,
\[
| \prob_{s \in \zo^{c \log n}}[C(g(s))=1] - \prob_{z \in \zo^n}[C(z)=1] | < 1/n^k.
\]

We note that assumption $H$  is realistic. Miltersen~\cite{mil:b:derandsurvey} has shown that it is implied by the following natural assumption, involving uniform complexity classes: for every $\epsilon >0$, there is a function $f \in E$ that cannot be computed in space $2^{\epsilon n}$ for infinitely many  lengths $n$.
 
 \section{Proofs}
 
 {\bf Proof of Theorem~\ref{t:ramseyuncond}.}
 \medskip
 
 Let us first review the probabilistic argument showing the existence of $2 \log n$-Ramsey graphs. A graph $G$ with $n$ vertices can be represented by a string of length ${n \choose 2}$. If we take at random such a graph and fix a subset of $k$ vertices, the probability that the set forms a clique or an independent set is $2^{-{k \choose 2}+1}$. The probability that this holds for some $k$-subset is bounded by
 \[
 \begin{array}{ll}
 {n \choose k} \cdot  2^{-{k \choose 2}+1} & \leq (\frac{en}{k})^k \cdot 2^{-{k \choose 2}+1} \\
 & = 2^{k \log \frac{en}{k} - {k \choose 2}+1}.
 
 \end{array}
 \]
 For $k = 2 \log n$, the above expression goes to $0$. Thus, for $n$ large enough, the probability that a graph $G$ is $2 \log n$-Ramsey is $\geq 0.99$.
 
 The key observation is that this argument remains valid if we take a distribution that is $2 \log^2 n$-wise independent. Thus, we can take a polynomial $p(X)$ of degree $2 \log^2 n$ over the field GF$[2^q]$, where $q = \log {n \choose 2}$. To the polynomial $p$ we associate the string
 $\tilde{p} = p(a_1)_1 \ldots p(a_{n \choose 2})_1$, where $a_1, \ldots, a_{n \choose 2}$ are the elements of the field and $(p(a))_1$ is the first bit of $p(a)$. When $p$ is random, this yields a distribution over strings of length ${n \choose 2}$ that is $2 \log^2 n$-wise independent. Observe that a polynomial $p$ is given by a string of length $\overline{n} = ( 2 \log^2 n +1) \log{n \choose 2} = O(\log^3 n)$. It follows that 
 \[
 \prob_{p \in \zo^{\overline{n}}} [ \tilde{p} \mbox{ is $ 2 \log n$-Ramsey}] \geq 0.99.
 \]
 In quasipolynomial time we can enumerate the graphs $\tilde{p}$, and 99\% of them are $2 \log n$-Ramsey.~\qed
\smallskip

\textbf{Note.}  By using an almost $k$-wise independent distribution (see~\cite{na-na:j:smallbias,al-gol-has-per:j:kwiseindep}), one can reduce the size of the list to $2^{O(\log^2 n)}$.
 \medskip
 
 {\bf Proof of Theorem~\ref{t:ramseycond} and of Theorem~\ref{t:ramseystrongcond}.}
 \medskip
 
 Let $p, \tilde{p}, \overline{n}$ be as in the proof of Theorem~\ref{t:ramseyuncond}. Thus:
 
\begin{itemize}
	\item $p \in \zo^{\overline{n}}$ represents a polynomial, 
	
	\item $\tilde{p}$ is built from the values taken by $p$ at all the elements of the underlying field, and represents a graph with $n$ vertices, 
	
	\item $\overline{n} = O(\log^3 n)$. 
 \end{itemize}
 Let us call a string $p$ \emph{good} if $\tilde{p}$ is a $2 \log n$-Ramsey graph.
 
 Checking that a string $p$ is not \emph{good} is an NP predicate. Indeed, $p$ is not \emph{good} iff $\exists (i_1, \ldots, i_{2 \log n}) \in [n]^{2 \log n}$ [ vertices $i_1, \ldots, i_{2 \log n}$ in $\tilde{p}$ form a clique or an independent set]. The $\exists$ is over a string of length polynomial in $|p|$ and the property in  the right parentheses can be checked by computing $O(\log^2 n)$ values of the polynomial $p$, which can be done in time polynomial in $|p|$.
 
 Assumption $H$ implies that there exists a pseudo-random generator $g: \zo^{c \log \overline{n}} \mapping \zo^{\overline{n}}$, computable in time polynomial in $\overline{n}$, that fools all NP predicates, and, in particular, also the one above.
 Since $99$\% of the $p$ are \emph{good}, it follows that for $90$\% of the seeds $s \in \zo^{c \log \overline{n}}$, $g(s)$ is \emph{good}, \ie, for $90$\% of $s$, $\widetilde{g(s)}$ is $2 \log n$-Ramsey. Note that from a seed $s$ we can compute $g(s)$ and next $\widetilde{g(s)}$ in time polynomial in $n$. If we do this for every seed $s \in \zo^{c \log \overline{n}}$, we obtain a list with $\overline{n}^c = O(\log^{3c} n)$ graphs of which at least $90$\% are $2 \log n$-Ramsey graphs.

Theorem~\ref{t:ramseystrongcond} is obtained by observing that $\{g(s) \mid s \in \zo^{c \log \overline{n}} \}$  is a list that can be computed in $\poly( \log n)$ time,  and most of its elements are concise representations of  $2 \log n$-Ramsey graphs.~\qed
 
 \section{Additional remarks}
 \label{s:remarks}
 \subsection{Constructing a single Ramsey graph from a list of graphs of which the majority are Ramsey graphs}
 
 M. Naor~\cite{mnaor:c:ramsey} has shown how to construct a Ramsey graph from a list of $m$ graphs such that all the graphs in the list, except at most $\alpha m$ of them, are $k$-Ramsey. We analyze what parameters are obtained, if we apply Naor's construction to the list of graphs in Theorem~\ref{t:ramseycond}.
 
 The main idea of Naor's construction is to use the product of two graphs $G_1 = (V_1, E_1)$ and $G_2 = (V_2, E_2)$, which is the graph whose set of vertices is $V_1 \times V_2$ and edges defined as follows: there is an edge between $(u_1,u_2)$ and $(v_1, v_2)$ if and only if $(u_1, v_1) \in E_1$ or $(u_1 = v_1) \mbox{ and } (u_2, v_2) \in E_2$. Then, one can observe that if $G_1$ is $k_1$-Ramsey and $G_2$ is $k_2$-Ramsey, the product graph, $G_1 \times G_2$ is $k_1 k_2$-Ramsey. Extending to the product of multiple graphs $G_1, G_2, \ldots, G_m$ where each $G_i$ is $k_i$-Ramsey, we obtain that the product graph is $k_1 k_2 \ldots k_m$-Ramsey.
 
 If we apply this construction to a list of $m$ graphs $G_1, G_2, \ldots, G_m$, each having $n$ vertices and such that $\prob_i[G_i \mbox{ is not } \mbox{$k$-Ramsey}] \leq \alpha$, we obtain that the product of $G_1, G_2, \ldots, G_m$ is a graph $G$ with $N = n^m$ vertices that is $t$-Ramsey for $t=n^{\alpha m} k^{(1-\alpha)m}$. For $\alpha \leq 1 /\log n$, we have $t \leq (2k)^m$. The list produced in Theorem~\ref{t:ramseycond} has $m = \log^c n$, $k = 2\log n$, and one can show that $\alpha \leq 1/\log n$. The product graph $G$ has $N=2^{\log^c n \log n}$  vertices and is $t$-Ramsey for $t \leq 2^{\log^c n \cdot \log \log n + O(1)} < 2^{(\log N)^{1 - \beta}}$, for some positive constant $\beta$. 
 
 Thus, under assumption $H$, there is a positive constant $\beta$ and a polynomial time algorithm that on input $1^N$ constructs a graph with $N$ vertices that is $2^{(\log N)^{1 - \beta}}$-Ramsey. Note that this is inferior to the parameters achieved by the unconditional construction of Barak, Rao, Shaltiel and Wigderson~\cite{barraoshawig:c:ramseygraph}.
 
 \subsection{Constructive lower bounds for the van der Waerden theorem}
 
 Van der Waerden Theorem is another classical result in Ramsey theory. It states that for every $c$ and $k$ there exists a number $n$ such that for any coloring of $\{1, \ldots, n\}$ with $c$ colors, there exists $k$ elements in arithmetic progression ($k$-AP) that have the same color. Let $W(c,k)$ be the smallest such $n$. One question is to find a constructive lower bound for $W(c,k)$. To simplify the discussion, let us focus on $W(2,k)$. 
 \smallskip
 
 In other words, the problem that we want to solve is the following: 
 \smallskip
 
 For any $k$, we want to find a value of $n= n(k)$ as large as possible and a  $2$-coloring of $\{1, \ldots, n\}$ such that no $k$-AP is monochromatic. Furthermore, we want the $2$-coloring to be computable in time polynomial in $n$.
 \smallskip
 
  Gasarch and Haeupler~\cite{gas-hae:j:waerden} have studied this problem. They present a probabilistic polynomial time construction for $n = \frac{2^{k-1}}{ek}-1$ (\ie, the $2$-coloring is obtained by a probabilistic algorithm running in $2^{O(k)}$ time) and a (deterministic) polynomial time construction for $n = \frac{2^{(k-1)(1-\epsilon)}}{4k}$ (\ie, the $2$-coloring is obtained in deterministic $2^{O(k/\epsilon)}$ time). Their constructions are based on the constructive version of the Lovasz Local Lemma due to Moser~\cite{mos:c:lovasz} and Moser and Tardos~\cite{mostar:j:lovasz}. The probabilistic algorithm of Gasarch and Haeupler is ``BPP-like", in the sense that it succeeds with probability $2/3$ and the correctness of the $2$-coloring produced by it can be checked in polynomial time. It follows that it can be derandomized under the hardness assumption that derandomizes BPP, using the Impagliazzo-Wigderson pseudo-random generator~\cite{imp-wig:c:pbpp}. It is
 interesting to remark that the new proof by Moser and Tardos of the Local Lovasz Lemma is essential here, because the success probability guaranteed by the classical proof is too small to be used in combination with the Impagliazzo-Wigderson pseudo-random generator.
 
 We proceed with the details.
 
 We use  the following hardness assumption $H'$ (weaker than assumption $H$), which is the one used to derandomize BPP~\cite{imp-wig:c:pbpp}.
 \medskip
 
 Assumption $H'$: There exists a function $f$ in $E$ such that, for some $\epsilon > 0$, for every $n$, $C_f(n) > 2^{\epsilon n}$.
\medskip

Impagliazzo and Wigderson~\cite{imp-wig:c:pbpp} have shown that, under assumption $H$, for every $k$, there is a constant $c$ and a pseudo-random generator $g:\zo^{c \log n} \mapping \zo^n$ that fools all $n^k$-size circuits and that is computable in time polynomial in $n$.

 \begin{proposition}
 \label{t:derandvdwaerden} Assume assumption $H'$. For every $k$, let $n= n(k) = \frac{2^{k-1}}{ek} - 1$. There exists a polynomial-time algorithm that on input $1^n$ $2$-colors the set $\{1, \ldots, n\}$ such that no $k$-AP is monochromatic.
 \end{proposition}
 $\proof$ The algorithm of Gasarch and Haeupler~\cite{gas-hae:j:waerden}, on input $1^n$, uses a random string $z$ of size $|z| = n^c$, for some constant $c$, and, with probability at least $2/3$, succeeds to $2$-color the set $\{1, \ldots, n\}$ such that no $k$-AP is monochromatic. Let us call a string $z$ to be \emph{good} for $n$ if the Gasarch-Haeupler algorithm on input $1^n$ and randomness $z$, produces a $2$-coloring with no monochromatic $k$-APs. Note that there exists a polynomial-time algorithm $A$ that checks if a string $z$ is \emph{good} or not, because the Gasarch-Haeupler algorithm runs in polynomial time and the number of $k$-APs inside $\{1, \ldots, n\}$ is bounded by $n^2/k$. Using assumption $H'$ and invoking the result of  Impagliazzo and Wigderson~\cite{imp-wig:c:pbpp}, we derive that there exists a constant $d$ and a pseudo-random generator $g: \zo^{d \log n} \mapping \zo^{n^c}$ such that 
 \[
 \prob_{s \in \zo^{d \log n}} [ A(g(s)) = \mbox{ good for $n$}] \geq 2/3 - 1/10 > 0.
 \] 
 Therefore if we try all possible seeds $s$ of length $d \log n$, we will find one $s$ such that $g(s)$ induces the Gasarch-Haeupler algorithm to $2$-color the set $\{1, \ldots, n\}$ such that no $k$-AP is monochromatic.~\qed
 \if01
 %
 %
 
 In this section, we show that a polynomial-time construction for $n = \frac{2^{k}}{8k}$ can be obtained almost directly from the Lovasz Local Lemma using an almost $k$-wise independent distribution. The Lovasz Local Lemma guarantees that a random $2$-coloring is good with probability $> 0$, and, since the distribution has a polynomially large support, we can do an exhaustive search through the elements in the support in polynomial time.
 
 We provide the details.
 
 A sequence $X_1, \ldots, X_n$ of random variables over $\zo$ is $\epsilon$-away from $k$-wise independence if for every $k$ distinct indeces $i_1, \ldots, i_k$ and for every $a \in \zo^k$,
 \[
| \prob[ X_{i_1} \ldots X_{i_k} = a] - 2^{-k}| \leq \epsilon. 
 \]
  Alon et al.~\cite{al-gol-has-per:j:kwiseindep} have shown that for every natural number $n$, every $\epsilon > 0$ and $k \leq n$, there exists a sequence $X_1, \ldots, X_n$ that is $\epsilon 2^{k/2}$-away from independent and such that the support of $X = (X_1, \ldots, X_n)$ has size $(n/\epsilon)^c$ and can be enumerated in time $(n/\epsilon)^c$, where $c$ is a constant.
  
  We also use the following particular case of the Lovasz Local Lemma (the so-called symmetric case).
  \begin{theorem}
  \label{t:symlll}
  Let ${\cal E}_1, \ldots, {\cal E}_m$ be events in a probability space. For all $i$, ${\cal E}_i$ satisfies $\prob[{\cal E}_i] \leq p$. Also each event is mutually independent of all other events except for at most $d$ and $ep(d+1) \leq 1$. Then
  \[
  \prob[\wedge_{i}\overline{{\cal E}_i} ] > 0.
  \]
  \end{theorem}
 We are now prepared to prove the announced result.
 \begin{theorem}
 There exists an algorithm running in time $2^{O(k)}$ that $2$-colors the set $\{1, \ldots, n\}$, where $n=\frac{2^{k}}{8k}$, such that no $k$-AP is monochromatic.

 \end{theorem}
 $\proof$ Let $\{T_1, \ldots, T_m\}$ be the set of $k$-APs inside $\{1, \ldots, n\}$. One element of $\{1, \ldots, n\}$ belongs to at most $\frac{nk}{k-1}$ such $k$-APs. The reason for this is that, given $x \in \{1, \ldots, n\}$, a $k$-AP containing $x$ is described by:
 \begin{itemize}
	\item the distance $\delta$ between consecutive terms. Note that $\delta \leq \frac{n}{k-1}$.
	\item the position of $x$ among the $k$ terms of the AP.
\end{itemize}
 This implies that one fixed $k$-AP intersects with at most $\frac{nk^2}{k-1}$ other $k$-APs. Also, $m$ is bounded by $\frac{n^2}{k-1}$.  
 Let 
 \[
 d = \frac{nk^2}{k-1}.
 \]
 If we randomly $2$-color the elements of $\{1, \ldots, n\}$, a fixed $k$-AP is monochromatic with probability $2^{-k+1}$. Note that this holds even if we choose the $2$-coloring from a distribution that is $k$-wise independent. Such distributions have supports that are still too large for our purposes. Therefore, we  choose the $2$-coloring from a distribution that is $2^{-k+1}$-away from $k$-independence, guaranteed by~\cite{al-gol-has-per:j:kwiseindep}. Such a distribution has size $(n/2^{-3k/2})^c$ and can be enumerated in time $(n/2^{-3k/2})^c$. With respect to this distribution the probability that a $k$-AP is monochromatic is bounded by 
 \[
 p = 2^{-k+1} + 2^{-k+1} = 2^{-k+2}.
 \]
 Let us call a $2$-coloring \emph{good}, if no $k$-AP is monochromatic. Note that checking if a $2$-coloring is \emph{good} can be done in time polynomial in $n$.
 For each $k$-AP $T_i$, we consider the event ${\cal E}_i$ that it is monochromatic. These events form a sequence of events satisfying the conditions in Theorem~\ref{t:symlll} with the above values of $p$ and $d$. The value of $n$ guarantees that $ep(d+1) \leq 1$. Thus the probability that there exists a \emph{good} $2$-coloring  is greater than $0$. Therefore, if we enumerate all the elements in the support of the distribution and check each of them if they are \emph{good} or not,  we will find a good $2$-coloring. ~\qed
 \fi
 \bibliography{c:/book-text/theory}

\bibliographystyle{alpha}
\end{document}